\documentstyle[eqsecnum,aps,multicol]{revtex}

\begin{document}
\draft
\preprint{}
\title{Mechanism of unconventional superconductivity induced by
skyrmion excitations in two-dimensional strongly-correlated electron
systems
}
\author{Takao Morinari}
\address{Yukawa Institute for Theoretical Physics, Kyoto University
Kyoto 606-8502, Japan
}
\date{\today}
\maketitle
\begin{abstract}
We propose a mechanism of unconventional superconductivity in
two-dimensional strongly-correlated electron systems.
We consider a two-dimensional Kondo lattice system or double-exchange
system with spin-orbit coupling arising from buckling of the plane.
We show that a Chern-Simons term is induced for a gauge field
describing the phase fluctuations of the localized spins.
Through the induced Chern-Simons term, carriers behave like
skyrmion excitations that lead to a destruction mechanism of magnetic
long-range order by carrier doping.
After magnetic long-range order is destroyed by carrier doping,
the Chern-Simons term plays a dominant role and the attractive
interaction between skyrmions leads to unconventional
superconductivity.
For the case of the ferromagnetic interaction between the localized
spins, the symmetry of the Cooper pair is p-wave ($p_x \pm
ip_y$). For the case of the antiferromagnetic interaction between the
localized spins, the symmetry of the Cooper pair is d-wave
($d_{x^2-y^2}$).
Applications to various systems are discussed, in particular to
the high-$T_c$ cuprates.
\end{abstract}

\begin{multicols}{2}

\narrowtext

\section{Introduction}
\label{sec_introduction}
Since the discovery of high-$T_c$ superconductivity in
 cuprates,\cite{BEDNORZ_MULLER}
a large number of studies have been invested to uncover its
mechanism of superconductivity.
Although the issue is still in controversy, there are some specific
properties concerning the mechanism of superconductivity.
First, the high-$T_c$ cuprates show two-dimensional highly anisotropic
behaviors.
From their structure, the high-$T_c$ cuprates have a layered
structure of $CuO_2$ planes with several $CuO_2$ layers 
sandwiched between insulating layers.
Reflecting this layered structure, measurements of the
resistivity\cite{ITO_ETAL} and optical
conductivity\cite{HOMES_ETAL,TAMASAKU_ETAL} show strong
anisotropic behaviors.
Furthermore, the electromagnetic properties of the superconducting
state is well described by a Josephson-coupled layer model.
\cite{KLEINER_ETAL,SHIBAUCHI_ETAL}
Secondly, it seems that there is a close relationship between
magnetism and superconductivity. 
In the absence of carriers, the system is a charge-transfer
insulator. \cite{FUJIMORI_ETAL}
Due to the large charge-transfer gap, a localized magnetic moment is 
produced at each Cu site.
These localized magnetic moments form antiferromagnetic long-range
order below N\'{e}el temperature. (N\'{e}el temperature is not equal
to zero because of weak inter-layer coupling.\cite{CHN})
When holes are doped in the $CuO_2$ plane, they occupy $O$
$2p_{\sigma}$ orbitals \cite{EELS,EELS2,XAS,XAS2} and destroy
antiferromagnetic long-range order.
As we increase the hole concentration, N\'{e}el temperature decreases.
Upon further doping, the antiferromagnetic long-range order is
destroyed and the system becomes the superconducting state.
Also in the phase of superconductivity, antiferromagnetic correlations
are observed.\cite{PENNINGTON,BIRGENEAU}

In contrast to conventional BCS superconductivity,
superconductivity in the high-$T_c$ cuprates is unconventional.
Symmetry of the Cooper pair is not s-wave but d-wave, or more
precisely, $d_{x^2-y^2}$-wave.\cite{D_WAVE} 
The fact that symmetry of the Cooper pair is d-wave,
superconductivity occurs in the vicinity of antiferromagnetic
long-range order, and the absence of the isotope effect\cite{ISOTOPE} 
suggests that the underlying mechanism of the high-$T_c$ cuprates be
ascribed to the antiferromagnetic correlations.

In addition, the structure of the $CuO_2$ plane seems to play an
important role for the pairing mechanism of high-$T_c$
superconductivity.
In $La_{2-x}Sr_xCuO_4$ system, suppression of
superconductivity is observed at a structural phase transition point
from an orthorhombic phase to a tetragonal phase. \cite{TAKAGI_ETAL}
Similar suppression of superconductivity, which is induced by the same 
kind of structural phase transition, is also observed in
$La_{2-x}Ba_xCuO_4$ system around $x=1/8$. \cite{MOODENBAUGH_ETAL}
Since there is buckling of the $CuO_2$ plane in the orthorhombic
phase, the fact that superconductivity occurs in the orthorhombic
phase suggests that buckling of the $CuO_2$ plane plays a significant
role for the occurrence of superconductivity.

In order to explain the mechanism of high-$T_c$ superconductivity in
the cuprates, a large number of theories have been proposed.
Among others, the spin-fluctuation theory \cite{MORIYA_UEDA} proposes a
pairing mechanism with d-wave symmetry, or $d_{x^2-y^2}$ symmetry.
For the d-p model, which is believed to capture the essential
properties of the $CuO_2$ plane, the spin-fluctuation theory predicts
$d_{x^2-y^2}$ pairing between d-orbital electrons at Cu
sites.\cite{TAKIMOTO,KOIKEGAMI}
The same kind of d-wave pairing, i.e., $d_{x^2-y^2}$ wave pairing, is
discussed in a different context.
In the resonating valence bond (RVB) theory
\cite{ANDERSON,NAGAOSA_LEE}, spinons form d-wave
pairing.\cite{KOTLIAR_LIU}
However, spinons are charge neutral quasiparticles
and the electric current is carried by holons in the RVB theory.
That is, d-wave pairing between spinons does not lead to d-wave
superconductivity by itself.
The d-wave pairing state of spinons rather describes the short-range
antiferromagnetic correlations in the phase without
antiferromagnetic long-range order.\cite{ANDERSON_ETAL}

Although the relationship between these two d-wave pairings has not
yet been clear,
the fact that the doped holes occupy Oxygen p-orbitals
\cite{EELS,EELS2,XAS,XAS2} and the sign of the Hall coefficient is 
positive \cite{HALL} indicates that carriers are doped holes.
We may say that it is doped holes to form the Cooper pair
in the superconducting state.
What we require to describe high-$T_c$ superconductivity is 
the d-wave pairing mechanism between doped holes.

In this paper, we propose such a mechanism of
superconductivity.
We consider a two-dimensional multi-band model which consists of a
carrier system and a localized spin system with strong coupling
between them.
As a typical candidate of such a system, we consider a Kondo lattice 
system or a double-exchange system. 
The high-$T_c$ cuprates can be described as a Kondo lattice
system\cite{IMADA_ETAL,MATSUKAWA_FUKUYAMA}, where carriers are doped holes and
localized spins are at Cu sites.
In order to take into account the effect of buckling of the plane,
we consider spin-orbit coupling arising from buckling.
We show that carriers induce frustration in the localized spin system
in the presence of spin-orbit coupling 
through Kondo or Hund coupling.
This frustration effect can be described in terms of skyrmion
excitations.
The skyrmion excitation is created at each position of the carriers
and plays a role of ``magnetic'' field for the carriers.
Because of the ``magnetic'' field produced around a carrier, the
Lorentz force acts on another carrier.
Due to this Lorentz force an attractive interaction is induced between 
carriers and leads to unconventional
superconductivity.\cite{MORINARI_D,MORINARI_P}

The outline of this paper is as follows.
In Sec.~\ref{sec_model}, we describe the model and the effect of
spin-orbit coupling.
In Sec.~\ref{sec_ferro}, we show the mechanism of superconductivity in 
the case of ferromagnetic interaction between localized spins.
Frustration effect induced by carriers is described as skyrmion
excitations through a Chern-Simons term for the gauge field which
describes the phase fluctuations of the localized spin system.
The fact that carriers behave like skyrmion excitations provides a
destruction mechanism of magnetic long-range order by carrier doping 
because the magnetic long-range order is destroyed by the skyrmion
excitations.
After magnetic long-range order is destroyed by the skyrmion
excitations, the Chern-Simons term plays a dominant role and 
the attractive interaction between skyrmions leads
to p-wave superconductivity.
In Sec.~\ref{sec_antiferro}, we show the mechanism of
superconductivity in the case of antiferromagnetic interaction between
localized spins.
In this case, the symmetry of the Cooper pair is d-wave, or more
precisely $d_{x^2-y^2}$ wave. 
We also show that the doping carrier induces
metal-insulator transition at the magnetic transition point.
In Sec.~\ref{sec_discussion}, we discuss applications to high-$T_c$
superconductivity and other systems.

\section{Model}
\label{sec_model}
We consider a two-dimensional multi-band model which can be reduced
 to a model consisting of carriers and localized spins with strong
 coupling between them.
Examples of such a model are the Kondo lattice system and the
 double-exchange system.
In order to include the buckling effect, we introduce spin-orbit
 coupling arising from it.

\subsection{Hamiltonian}
The Hamiltonian of the Kondo-lattice system or the double-exchange
 system with the spin-orbit coupling term may be written in the
 following form:
\begin{equation}
H = H_0 + H_{\rm int} + H_{\rm so} + H_{\rm spin}.
\label{hamiltonian}
\end{equation}
Here the first term is the kinetic energy term for the carrier system,
\begin{equation}
H_0 = -t \sum_{\langle i,j \rangle} \left( c_i^{\dagger}
c_j + H.c. \right),
\end{equation}
where the summation is taken over the nearest neighbor sites. 
Carrier operators are represented by a spinor notation,
\begin{equation}
c_i^{\dagger}=
\left(
\begin{array}{cc}
c_{i\uparrow}^{\dagger} & c_{i\downarrow}^{\dagger}
\end{array}
\right),
\hspace{1em}
c_j=\left(
\begin{array}{c}
c_{j\uparrow} \\
c_{j\downarrow}
\end{array}
\right).
\end{equation}
The second term $H_{\rm int}$ represents Kondo or Hund coupling
between
the carrier spin ${\bf s}_j$ and the localized spin ${\bf S}_j$:
\begin{equation}
H_{\rm int}=-J_c \sum_j {\bf s}_j \cdot {\bf S}_j,
\end{equation}
where ${\bf s}_j = \frac12 c_j^{\dagger} 
{\mbox{\boldmath ${\bf \sigma}$}} c_j$ with
the components of ${\mbox{\boldmath ${\bf \sigma}$}} = 
(\sigma_1,\sigma_2,\sigma_3)$ being the Pauli spin matrices.
We take $1$, $2$, and $3$ for the axes in spin space.
We assume that $|J_c|$ is the largest energy scale in the Hamiltonian
(\ref{hamiltonian}).

The third term $H_{\rm so}$ in Eq.~(\ref{hamiltonian}) 
represents spin-orbit coupling arising from 
buckling,
\begin{equation}
H_{\rm so} = i \sum_{j} 
\sum_{{\mbox{\boldmath ${\bf \eta}$}}=(a,0),(0,a)} 
c_j^{\dagger} 
{\mbox{\boldmath ${\bf \lambda}$}}^{({\mbox{\boldmath ${\bf \eta}$}})}
\cdot {\mbox{\boldmath ${\bf \sigma}$}} 
c_{j+ {\mbox{\boldmath ${\bf \eta}$}}} + H.c.,
\end{equation}
where $a$ is the lattice constant and the vectors
${\mbox{\boldmath ${\bf \lambda}$}}^{({\mbox{\boldmath ${\bf
\eta}$}})} = (\lambda_1^{({\mbox{\boldmath ${\bf \eta}$}})},
\lambda_2^{({\mbox{\boldmath ${\bf \eta}$}})}, 0)$
are proportional to both spin-orbit coupling of ions and the angle
of buckling.
The simplest example for the spin-orbit coupling term is presented in
the appendix of Ref.~\cite{ANDO} in which the spin-orbit coupling term 
for the s and p orbitals is derived.
The spin-orbit coupling terms of the high-$T_c$ cuprates are shown in
Refs.~\cite{BONESTEEL_ETAL,SCHEKHTMAN_ETAL,KOSHIBAE_ETAL}.
For simplicity, we assume the vectors 
${\mbox{\boldmath ${\bf \lambda}$}}^{({\mbox{\boldmath ${\bf
\eta}$}})}$
for the orthorhombic phase of $YBa_2Cu_3O_{7-\delta}$, that is, 
${\mbox{\boldmath ${\bf \lambda}$}}^{(-a,0)} = 
{\mbox{\boldmath ${\bf \lambda}$}}^{(a,0)}$
and 
${\mbox{\boldmath ${\bf \lambda}$}}^{(0,-a)} = 
{\mbox{\boldmath ${\bf \lambda}$}}^{(0,a)}$
with 
$|{\mbox{\boldmath ${\bf \lambda}$}}^{(a,0)}| = 
|{\mbox{\boldmath ${\bf \lambda}$}}^{(0,a)}| \equiv \lambda$.

The last term $H_{\rm spin}$ in Eq.~(\ref{hamiltonian}) represents the
interaction between the localized spins.
For $H_{\rm spin}$ we take the Heisenberg Hamiltonian
\begin{equation}
H_{\rm spin} = J \sum_{\langle i,j \rangle} {\bf S}_i \cdot {\bf S}_j.
\label{eq_spin}
\end{equation}
In general, the Dzyaloshinskii-Moriya interaction is induced between
the localized spins when there is spin-orbit coupling like $H_{\rm
so}$.
However, the Dzyaloshinskii-Moriya interaction does not play an
important role in our mechanism of superconductivity.
We neglect the Dzyaloshinskii-Moriya interaction in the following
analysis.

\subsection{Effect of spin-orbit coupling}
\label{sec_so}
In the last subsection, we have introduced spin-orbit coupling arising
from buckling of the plane.
Since this spin-orbit coupling plays an essential role in our
mechanism of superconductivity, first we need to discuss
the effect of spin-orbit coupling.

The effect of the spin-orbit coupling term $H_{\rm so}$ becomes
apparent when we combine $H_{\rm so}$ with the kinetic energy term
$H_0$:
\begin{eqnarray}
H_0 + H_{\rm so} 
&=&
\sum_i 
\sum_{\mbox{\boldmath ${\bf \eta}$}}
\left[ c_i^{\dagger}
\left(
-t \sigma_0 + i 
{\mbox{\boldmath ${\bf \lambda}$}}^{({\mbox{\boldmath ${\bf \eta}$}})}
\cdot {\mbox{\boldmath ${\bf \sigma}$}}
\right) c_{i+{\mbox{\boldmath ${\bf \eta}$}}}
+ {\rm H.c.} \right] \nonumber \\
&=&
- \sqrt{t^2 + \lambda^2} 
\sum_i \sum_{{\mbox{\boldmath ${\bf \eta}$}}}
\left[ c_i^{\dagger}
\exp \left( - \frac{i}{t}
{\mbox{\boldmath ${\bf \lambda}$}}^{({\mbox{\boldmath ${\bf \eta}$}})}
\cdot {\mbox{\boldmath ${\bf \sigma}$}} \right) \right. \nonumber \\
& & \left. \times
c_{i+{\mbox{\boldmath ${\bf \eta}$}}} + {\rm H.c.} \right],
\end{eqnarray}
up to $O((\lambda/t)^2)$ in the exponential,
where $\sigma_0$ is the unit matrix in spin space.
The factor $\exp \left( - \frac{i}{t}
{\mbox{\boldmath ${\bf \lambda}$}}^{({\mbox{\boldmath ${\bf \eta}$}})}
\cdot {\mbox{\boldmath ${\bf \sigma}$}} \right)$ has the form of  a
unitary transformation of rotation in spin space. 
The axis of rotation is parallel to the vector 
${\mbox{\boldmath ${\bf \lambda}$}}^{({\mbox{\boldmath ${\bf
\eta}$}})}$ and the angle of rotation is $2\lambda/t$.

The presence of this rotation at every hopping process of the
carriers implies that the carriers introduce disorder in the localized
spin system through strong coupling, $H_{\rm int}$ between the 
carriers and the localized spins.
Disorder produced by the carrier hopping processes provides a
destruction mechanism of magnetic long-range order in the localized
spin system.
In Secs.~\ref{sec_skyF} and \ref{sec_skyAF} 
we will show that this destruction
mechanism of magnetic long-range order is represented as the effect of
skyrmion excitations.

\section{Ferromagnetic case}
\label{sec_ferro}
In order to illustrate the mechanism of superconductivity, 
we first consider the case of ferromagnetic interaction between the
localized spins, that is, $J<0$ in $H_{\rm spin}$.

\subsection{Schwinger bosons}
\label{sec_sb}
We are interested in the mechanism of superconductivity  based on the
fluctuations of the localized spins.
In order to describe the localized spins,
we introduce Schwinger bosons.
Description of the localized spin system in terms of the Schwinger
bosons has some advantages.
First of all, it is straightforward to describe the magnetic
long-range ordered state.
The phase with magnetic long-range order is described by Bose-Einstein
condensation of Schwinger bosons.\cite{AROVAS_AUERBACH,AUERBACH}
Another advantage is that we can directly construct rotation
matrices for carrier's spins.
Such matrices turn out to be useful for the description of the
localized spin fluctuation effect on the carriers.

Each localized spin can be described by Schwinger bosons,
\begin{equation}
{\bf S}_j = \frac12 
\left(
\begin{array}{cc}
z_{j\uparrow}^{\dagger} & z_{j\downarrow}^{\dagger}
\end{array}
\right)
{\mbox{\boldmath ${\bf \sigma}$}} 
\left(
\begin{array}{c}
z_{j\uparrow} \\
 z_{j\downarrow}
\end{array}
\right).
\end{equation}
Here $z_{j\sigma}^{\dagger}$ and $z_{j\sigma}$ are Schwinger bosons at 
site $j$
and obey boson commutation relations: 
$\left[ z_{i\sigma},z_{j\sigma'}^{\dagger} \right] = \delta_{ij}
\delta_{\sigma \sigma'}$ and
$\left[ z_{i\sigma},z_{j\sigma'}\right] = 
\left[ z_{i\sigma}^{\dagger},z_{j\sigma'}^{\dagger} \right] = 0$
and the constraint 
$\sum_{\sigma} z_{j\sigma}^{\dagger} z_{j\sigma}=2S$.
\cite{AROVAS_AUERBACH,AUERBACH}
In the following, we consider the case of $S=1/2$ for simplicity.
However,
it is straightforward to extend the following arguments to general
values of $S$.

In terms of Schwinger boson fields, the Hamltonian for the localized
spin system can be written, up to constant term, in the following
form:
\begin{equation}
H_{\rm spin} = - \frac12 |J| \sum_{\langle i,j \rangle}
F_{ij}^{\dagger} F_{ij},
\label{eq_ff}
\end{equation}
where $F_{ij} = \sum_{\sigma} z_{i\sigma}^{\dagger} z_{j\sigma}$.
Turning to the path-integral formalism, we introduce a
Stratonovich-Hubbard
field $Q_{ij}$ and $\overline{Q}_{ij}$ to decouple the interaction
term $F_{ij}^{\dagger} F_{ij}$ :
\begin{equation}
{\cal Z}_{\rm spin} = \int {\cal D} \overline{z} {\cal D} z 
{\cal D} \lambda^{\rm SB} {\cal D} \overline{Q} {\cal D} Q 
\exp \left( -S_{\rm spin} \right),
\end{equation}
where
\begin{eqnarray}
S_{\rm spin} 
&=& \int_0^{\beta} d\tau \left\{
\sum_{j\sigma} \overline{z}_{j\sigma} 
\left( \partial_{\tau} - i \lambda_j^{\rm SB} \right)
z_{j\sigma} 
\right. \nonumber \\ & & \left. 
+\frac{|J|}{2} \sum_{\langle i,j \rangle}
\left[ ~\overline{Q}_{ij} Q_{ij} 
- \sum_{\sigma} \left( Q_{ij} \overline{z}_{j\sigma} z_{i\sigma}
+ \overline{Q}_{ij} \overline{z}_{i\sigma} z_{j\sigma} \right) \right]
\right\},\nonumber \\
\end{eqnarray}
where the $\tau$ dependence of all fields is implicit and 
$\lambda^{\rm SB}_j$ is introduced to impose the constraint 
$\sum_{\sigma} \overline{z}_{j\sigma} z_{j\sigma} = 1$.

Now let us study the localized spin fluctuations.
The spin fluctuations are represented by $Q_{ij}$ because we obtain
$\langle Q_{ij} \rangle = \sum_{\sigma} 
\langle \overline{z}_{i\sigma} z_{j\sigma}\rangle$
at the saddle point.
The spin fluctuation $Q_{ij}$ consists of the phase fluctuations and
the amplitude fluctuations.
Since the latter turns out to be a high-energy mode,
we focus on the phase fluctuations of $Q_{ij}$.

The phase fluctuations of $Q_{ij}$ are connected with the local gauge
transformation of 
$\overline{z}_{j\sigma}$
(or $z_{j\sigma}^{\dagger}$)
and $z_{j\sigma}$ at each
site.
In fact, Eq.~(\ref{eq_ff}) is invariant under the local gauge
transformation
$z_{j\sigma} \rightarrow z_{j\sigma} \exp (-i\theta_j )$.
In the action $S_{\rm spin}$, this gauge transformation involves
a transformation in the phase of $Q_{ij}$.
That is, the transformation in the phase of $Q_{ij}$ can be described
by a gauge field.
Introducing a gauge field and the mean amplitude 
$Q=\langle Q_{ij} \rangle$ and taking a continuum limit,
we may write the action $S_{\rm spin}$ in the following form:
\begin{eqnarray}
S_{\rm spin} 
&=& \int_0^{\beta} \!\!\! d\tau \int\!\!\! d^2 {\bf r}
\sum_{\sigma} \left[
\overline{z}_{\sigma} ({\bf r},\tau)
\left( \partial_{\tau} - i {\cal A}_{\tau}^{\rm SB} 
- i\lambda^{\rm SB} \right) z_{\sigma} ({\bf r},\tau) 
\right. \nonumber \\
& & \left. 
+ \frac{|J| Q}{2} \left| \left( \nabla 
 - i {\mbox{\boldmath ${\cal A}$}}^{\rm SB} \right) 
 z_{\sigma} ({\bf r},\tau) \right|^2 
\right],
\label{eq_spin_cont}
\end{eqnarray}
where 
\begin{equation}
{\cal A}^{\rm SB}_{\mu} ({\bf r},\tau) = -i 
\sum_{\sigma} \overline{z}_{\sigma} ({\bf r},\tau)
\partial_{\mu} z_{\sigma} ({\bf r},\tau).
\end{equation}
Note that Eq.~(\ref{eq_spin_cont}) is invariant under the gauge
transformation:
$z_{\sigma} ({\bf r},\tau) \rightarrow z_{\sigma} ({\bf r},\tau) \exp
\left( -i\theta ({\bf r},\tau) \right)$, and 
${\cal A}_{\mu}^{\rm SB} ({\bf r},\tau) \rightarrow {\cal
A}_{\mu}^{\rm SB} ({\bf
r},\tau) - \partial_{\mu} \theta ({\bf r},\tau)$.

\subsection{Gauge field description of the strong correlations}
The spin fluctuations of the localized spin system affect the carrier
system through strong coupling $H_{\rm int}$.
We may say that this strong coupling between the carriers and the
localized spins is the origin of strong correlations.
In order to take into account this strong correlation effect
we rotate the spin of the carrier so as to be in the direction
of the localized spin at the same site.
Through this transformation, the effect of the spin fluctuations
on the carrier system is described by coupling to a gauge field.

The action of the carrier system with $H_{\rm int}$ is 
\begin{equation}
S_c + S_{\rm int} = \int_0^{\beta} d \tau 
\left[
\sum_j 
\overline{c}_j(\tau) (\partial_{\tau}-\mu) c_j(\tau)
+ H_c + H_{\rm int} \right],
\end{equation}
where
\begin{eqnarray}
H_c + H_{\rm int}
&=& -\sqrt{t^2+\lambda^2}
\sum_j \sum_{{\mbox{\boldmath ${\bf \eta}$}}}
\left[
\overline{c}_{j+{\mbox{\boldmath ${\bf \eta}$}}} (\tau) 
{\rm e}^{-\frac{i}{t} 
{\mbox{\boldmath ${\bf \lambda}$}}^{({\mbox{\boldmath ${\bf \eta}$}})}
\cdot {\mbox{\boldmath ${\bf \sigma}$}}}
c_j (\tau) \right. \nonumber \\ & & \left.
+ \overline{c}_{j} (\tau)  
{\rm e}^{ \frac{i}{t} 
{\mbox{\boldmath ${\bf \lambda}$}}^{({\mbox{\boldmath ${\bf \eta}$}})}
\cdot {\mbox{\boldmath ${\bf \sigma}$}}}
c_{j+{\mbox{\boldmath ${\bf \eta}$}}} (\tau)
\right] \nonumber \\
& & 
- J_c \sum_j {\bf s}_j (\tau ) \cdot {\bf S}_j (\tau ).
\end{eqnarray}
In order to rotate the carrier's spin ${\bf s}_j$ in the direction 
of the localized spin ${\bf S}_j$, 
we perform the following unitary transformation:
\begin{equation}
c_j \rightarrow U_j {c}_j, \hspace{1em}
\overline{c}_j \rightarrow \overline{c}_j \overline{U}_j,
\end{equation}
where
\begin{equation}
U_j = \left(
\begin{array}{cc}
z_{j\uparrow} & -\overline{z}_{j\downarrow} \\
z_{j\downarrow} & \overline{z}_{j\uparrow} 
\end{array}
\right),\hspace{1em}
\overline{U}_j = \left(
\begin{array}{cc}
\overline{z}_{j\uparrow} & \overline{z}_{j\downarrow} \\
- z_{j\downarrow} & z_{j\uparrow} 
\end{array}
\right).
\label{eq_defU}
\end{equation}
Under these transformations, $H_{\rm int}$ is reduced to
$H_{\rm int} 
\rightarrow
H_{\rm int} = - \frac{J_c}{4} \sum_j 
\overline{c}_j \sigma_z {c}_j$.
In the hopping term, the following phase factor is introduced:
\begin{equation}
\overline{U}_{j+{\mbox{\boldmath ${\bf \eta}$}}} U_j
=
\exp \left( -i 
{\mbox{\boldmath ${\bf \eta}$}} \cdot 
{\mbox{\boldmath ${\bf {\cal A}}$}}_{j{\mbox{\boldmath ${\bf \eta}$}}}
\right).
\end{equation}
If the phase fluctuation 
${\mbox{\boldmath ${\bf {\cal A}}$}}_{j{\mbox{\boldmath ${\bf
\eta}$}}}$ is suffciently slowly varying, 
we can take the continuum limit. (We will discuss the validity of this 
approximation in Sec.~\ref{sec_condition}.)
Thus we obtain
\begin{equation}
S_c + S_{\rm int} = \int_0^{\beta} d\tau \int d^2 {\bf r}
\overline{\psi} ({\bf r},\tau) 
G^{-1} (\{ {\hat k}_{\mu} + {\cal A}_{\mu} \})
\psi ({\bf r},\tau),
\label{eq_cint}
\end{equation}
where ${\hat k}_{\mu}=-i\partial_{\mu}$ and ${\cal A}_{\mu}$ is the
$SU(2)$ gauge field arising from the spin fluctuations of the localized 
spins,
\begin{equation}
{\cal A}_{\mu}
=\sum_{a=1,2,3} {\cal A}_{\mu}^a \sigma_a 
= -i \overline{U} \partial_{\mu} U.
\label{eq_A}
\end{equation}
Note that ${\cal A}^3_{\mu}={\cal A}_{\mu}^{\rm SB}$.
The inverse of Green's function is
\begin{equation}
G^{-1} (\{ k_{\mu} \})
= 
\left( ik_{\tau} + \xi_k
\right) \sigma_0
+ {\bf g} ({\bf k}) 
\cdot {\mbox{\boldmath ${\bf \sigma}$}},
\label{eq_Ginv}
\end{equation}
with $\xi_k=k^2/(2m)-\mu$ ($1/(2m) \equiv t$).
Here ${\bf g}({\bf k})$ is given by
\begin{equation}
{\bf g} ({\bf k}) 
= 
2 {\mbox{\boldmath ${\bf \lambda}$}}^{(a,0)} k_x
+2 {\mbox{\boldmath ${\bf \lambda}$}}^{(0,a)} k_y
-\frac{J_c}{4} {\hat e}_3
\label{eq_g}
\end{equation}
As a result, the total action may be written in the following form:
\begin{eqnarray}
S &=& S_c + S_{\rm int} + S_{\rm spin} \nonumber \\
&=& \int_0^{\beta} d\tau \int d^2 {\bf r}
\left\{
\overline{\psi} ({\bf r},\tau) 
G^{-1} (\{ {\hat k}_{\mu} + {\cal A}_{\mu} \})
\psi ({\bf r},\tau) 
\right. \nonumber \\ & & \left. 
+ \sum_{\sigma} \left[ 
\overline{z}_{\sigma} ({\bf r},\tau)
\left( \partial_{\tau} - i {\cal A}_{\tau}^3 - i\lambda^{\rm SB}
\right)
z_{\sigma} ({\bf r},\tau) 
\right. \right. \nonumber \\
& & \left. \left. + \frac{|J| Q}{2} |\left( \nabla 
  - i {\mbox{\boldmath ${\cal A}$}}^3 \right) 
       z_{\sigma} ({\bf r},\tau)|^2  \right]
\right\}
\label{eq_actionF}
\end{eqnarray}

\subsection{Effective action of the gauge field}
From Eq.~(\ref{eq_actionF}) we can see that the fluctuations of the
localized spins affect the carrier system through the gauge field
${\cal A}_{\mu}$.
Therefore, in order to investigate the effect of the spin fluctuations 
on the carrier system, we need to study the properties of the gauge
field ${\cal A}_{\mu}$,
that is, we need to calculate the effective action of the gauge field 
${\cal A}_{\mu}$.
The effective action $S_{\cal A}$,
consists of two parts:
\begin{equation}
S_{\cal A}=S_{\cal A}^{\rm c}+S_{\cal A}^{\rm spin}.
\end{equation}
Here 
$S_{\cal A}^c$ is the contribution from the carrier system
and $S_{\cal A}^{\rm spin}$ is that from the localized spin system.

We obtain $S_{\cal A}^c$ by integrating out the carrier fields.
From Eqs. (\ref{eq_Ginv}) and (\ref{eq_g}), one can see that 
${\mbox{\boldmath ${\bf \lambda}$}}^{({\mbox{\boldmath ${\bf
\eta}$}})}$ play a role of the Dirac $\gamma$ matrices in $2+1$
dimension and $J_c$ plays a role of the Dirac fermion mass.
The derivation of the effective action of the gauge field ${\cal
A}_{\mu}$ is similar to that for massive Dirac fermions in $2+1$
dimension. \cite{CS_TERM,HST,BDP,READ_GREEN}
We find that the Chern-Simons term for the gauge field ${\cal
A}_{\mu}$ is induced.
(Detail of the calculation is presented in the Appendix.)
The induced Chern-Simons term has the following form:
\begin{equation}
S_{\cal A}^{\rm c} = \frac{i\theta}{2\pi} \int_0^{\beta} d\tau \int
d^2 {\bf r} {\cal A}_{\tau}^3 \left( \partial_x A_y^3-\partial_y A_x^3
\right).
\label{eq_cs}
\end{equation}
Here we retain only the third component of the $SU(2)$ gauge field
because it describes the ferromagnetic spin fluctuations of the
localized spins.
The coefficient of the Chern-Simons term is
\begin{equation}
\theta=\frac12 ~{\rm sgn} \left( J_c \Lambda \right),
\label{eq_c}
\end{equation}
with $\Lambda = \lambda_1^{(a,0)} \lambda_2^{(0,a)} 
- \lambda_2^{(a,0)} \lambda_1^{(0,a)}$.
Equation (\ref{eq_c}) is the expression at zero temperature.
For finite temperature, Eq.~(\ref{eq_c}) is slightly modified. 
However, we can neglect finite temperature effect as long as 
$\beta |J_c| \gg 1$.

The action (\ref{eq_cs}) represents the combined effect of the
spin-orbit coupling term $H_{\rm so}$ and the strong coupling term
$H_{\rm int}$.
This effect is qualitatively described in Sec.~\ref{sec_so}.
That is, destruction of magnetic long-range order.
In the gauge field description,
the effect is described by the Chern-Simons term for the gauge field.

For the contribution from the localized spin system $S_{\cal A}^{\rm
spin}$, it depends on whether there is magnetic long-range order or not.
In the absence of magnetic long-range order, 
$S_{\cal A}^{\rm spin}$ may have a form of the Maxwell term.
Since there is an extra derivative in the Maxwell term compared with
the Chern-Simons term, we expect that the Maxwell term has unimportant
effect for the long-wavelength and low-energy theory.
On the other hand,
in the presence of magnetic long-range order, 
the gauge field ${\cal A}_{\mu}^3 (={\cal A}_{\mu}^{\rm SB})$ 
becomes massive since Schwinger bosons form Bose-Einstein condensate.
(That is, Schwinger bosons are in the Meissner phase with respect to
the gauge field ${\cal A}_{\mu}^3$.)

\subsection{Skyrmion excitations}
\label{sec_skyF}
Since coupling between the carriers and the localized spins is made
only through the gauge field ${\cal A}_{\mu}^3$,
we may write the effective action of the carrier system in the
following form:
\begin{eqnarray}
S_c^{\rm eff}
&=&
\int_0^{\beta} d \tau \int d^2 {\bf r}
\overline{\psi} ({\bf r},\tau) \nonumber \\
& & \times \left[ \partial_{\tau} -\mu + i {\cal A}_{\tau}^3 
+ \frac{1}{2m} \left( -i\nabla + 
{\mbox{\boldmath ${\cal A}$}}^3 \right)^2 \right] \psi ({\bf r},\tau)
\nonumber \\
& & + \frac{i\theta}{2\pi} \int_0^{\beta} d\tau \int
d^2 {\bf r} {\cal A}_{\tau}^3 \left( \partial_x A_y^3-\partial_y A_x^3
\right) + S_{\cal A}^{\rm spin}.
\label{eq_ceff}
\end{eqnarray}
From this action, one can derive an important relationship between the 
carriers and excitations in the localized spin system.
The variation of $S_c^{\rm eff}$ with respect to ${\cal A}_{\tau}^3$
yields
\begin{equation}
\sum_{\sigma} s_{\sigma} \rho_{\sigma} ({\bf r},\tau)
= -\frac{\theta}{2\pi} \left(
\partial_x {\cal A}_y^3 - \partial_y {\cal A}_x^3 \right),
\label{eq_sk}
\end{equation}
where $s_{\uparrow}=+1$ and $s_{\downarrow}=-1$.
This equation implies that a gauge flux is produced at the position of
the carrier.
This gauge flux corresponds to the skyrmion excitation similar to a
topological excitation \cite{RAJARAMAN} of the non-linear sigma model
or $CP^1$ model
\cite{SKYRMION1,SKYRMION2,ALM,SKYRMION3,SKYRMION4,SKYRMION5}

The gauge fluxes produced by each carrier play a role of vortices
introduced in a BCS superconductor.
In a BCS superconductor, which is Bose-Einstein condensation 
of Cooper pairs, disorder is introduced by vortices, 
or the electromagnetic gauge fluxes.
As mentioned in Sec.~\ref{sec_sb}, the magnetic long-range order
in the localized spin system is described by Bose-Einstein
condensation of Schwinger bosons.
Since the Schwinger bosons couple to the gauge field ${\cal
A}_{\mu}^3$, its gauge fluxes introduce disorder in Bose-Einstein
condensation of Schwinger bosons.
This implies that disorder is introduced by skyrmion excitations, or
carrier doping through the relationship (\ref{eq_sk}).
In the presence of magnetic long-range order,
the skyrmion excitations have an excitation energy gap.
After magnetic long-range order is destroyed by skyrmion excitations, 
they become gapless excitations and the Chern-Simons term plays a
dominant role in long-wavelength and low-energy physics.

Before moving on to study the interaction between these skyrmion
excitations, we make some comments on the time-reversal symmetry.
From the coefficient of the Chern-Simons term Eq.~(\ref{eq_c}), we see 
that the Berry phase induced by the Chern-Simons gauge flux is $2\pi$.
Such a Berry phase preserves the statistics of particles.
By contrast, in the anyon system \cite{ANYON1,ANYON2} 
the Berry phase is $\pi/2$.
The Berry phase of non-integer multiples of $\pi$ implies the
time-reversal symmetry breaking,\cite{KIVELSON_ROKHSAR}
whereas in our case, there is no such implication of the
time-reversal
symmetry breaking arising from the Berry phase.
In addition, there is no mean ``magnetic'' field as long as 
$\langle \sum_{\sigma} s_{\sigma} \rho_{\sigma} ({\bf r},\tau) \rangle 
= 0$.
On the other hand, since the gauge field describes the localized spin
fluctuations, both the time-reversal symmetry and the parity are
broken in the localized spin system by the presence of the
Chern-Simons term.

\subsection{Superconductivity}
\label{sec_sc}
In the phase without magnetic long-range order in the localized spin
system, the Chern-Simons term plays a dominant role in the action
$S_{\cal A}$.
In this phase, an attractive interaction is induced between skyrmions.
Through this attractive interaction, carriers form the Cooper pair.

Intuitively, we can understand the occurrence of an attractive
interaction between skyrmions as follows.
As discussed in Sec.~\ref{sec_so}, the carrier rotates its spin at
each hopping process due to the spin orbit coupling term $H_{\rm so}$.
These rotations of the carrier's spin affects the localized spins
through the strong coupling term $H_{\rm int}$.
This effect can be described as the formation of a spin configuration
in the localized spin system.
This spin configuration carried by each carrier generates a
``magnetic'' field around other carriers through the Berry phase.
Therefore when a carrier passes another carrier with velocity, a 
Lorentz force acts between them.
This Lorentz force plays a role of pairing interaction.

Now let us go into detail.
The Lorentz force is derived from minimal coupling between the
carriers and the gauge field ${\cal A}^3_{\mu}$:
\begin{equation}
V_{\rm int}=\int d^2 {\bf r} \frac{1}{2m} \sum_{\sigma} s_{\sigma}
{\psi}_{\sigma}^{\dagger} ({\bf r})
\left( 
  \hat{{\bf k}}\cdot {\mbox{\boldmath ${\cal A}$}}^3 +
  {\mbox{\boldmath ${\cal A}$}}^3 \cdot \hat{{\bf k}}
\right)
\psi_{\sigma} ({\bf r}).
\label{eq_int}
\end{equation}
From Eqs. (\ref{eq_sk}) and (\ref{eq_int}), we eliminate the gauge
field ${\mbox{\boldmath ${\bf {\cal A}}$}}^3$ upon taking the Coulomb
gauge.
Thus we obtain
\begin{eqnarray}
V_{\rm int} 
&=& \frac{1}{2\Omega} 
\sum_{{\bf k}_1 \neq {\bf k}_2, {\bf q}}
s_{\sigma_1} s_{\sigma_2}
V_{{\bf k}_1 {\bf k}_2} \nonumber \\
& & \times c_{{\bf k}_1+{\bf q}/2, \sigma_1}^{\dagger}
c_{-{\bf k}_1+{\bf q}/2, \sigma_2}^{\dagger} c_{-{\bf k}_2+{\bf q}/2,
\sigma_2}
c_{{\bf k}_2+{\bf q}/2, \sigma_1}, 
\label{eq_int2} 
\end{eqnarray}
where $\Omega$ is the area of the system and
\begin{equation}
V_{{\bf k}_1 {\bf k}_2}
= \frac{4\pi i}{m \theta}
\frac{{\bf k}_1 \times {\bf k}_2}{|{\bf k}_1-{\bf k}_2|^2}.
\end{equation}
Equation (\ref{eq_int2}) represents the interaction between the
carriers mediated by the gauge field ${\cal A}_{\mu}^3$.
Note that in this equation no parameters 
characterizing the skyrmion excitations appears except for the
Chern-Simons term coefficient $\theta$.
Furthermore, there is no retardation effect.

Now we may write the effective Hamiltonian for the carrier system 
in the following form:
\begin{eqnarray}
H &=& \sum_{{\bf k} \sigma} \xi_k 
c_{{\bf k}\sigma}^{\dagger} c_{{\bf k}\sigma} + \frac{1}{2\Omega}
\sum_{{\bf k}_1 \neq {\bf k}_2}
\sum_{\sigma_1 \sigma_2} s_{\sigma_1} s_{\sigma_2}
V_{{\bf k}_1 {\bf k}_2} 
\nonumber \\ 
& & \times 
c_{{\bf k}_1 \sigma_1}^{\dagger}
c_{-{\bf k}_1 \sigma_2}^{\dagger}
c_{-{\bf k}_2 \sigma_2}
c_{{\bf k}_2 \sigma_1}.
\label{eq_hamiltonian}
\end{eqnarray}
Here we set ${\bf q}={\bf 0}$ in the interaction term to focus on the
equilibrium state.
We investigate the possibility of superconductivity based on the
Hamiltonian (\ref{eq_hamiltonian}).
For simplicity, we analyze the Hamiltonian within a mean field theory.
We introduce the following pairing matrices:
\begin{equation}
\Delta^{\bf k}_{\sigma_1\sigma_2}
= \frac{1}{\Omega} \sum_{{\bf k}'(\neq {\bf k})}
V_{{\bf k}{\bf k}'}
\langle 
c_{-{\bf k}'\sigma_2}
c_{{\bf k}' \sigma_1}
\rangle,
\end{equation}
\begin{equation}
\left( \Delta^{\bf k} \right)^{\dagger}_{\sigma_1 \sigma_2}
= \frac{1}{\Omega} \sum_{{\bf k}'(\neq {\bf k})}
V_{{\bf k}'{\bf k}}
\langle 
c_{{\bf k}'\sigma_1}^{\dagger}
c_{-{\bf k}'\sigma_2}^{\dagger}
\rangle.
\end{equation}
In terms of these matrices, we define the mean field Hamiltonian
\begin{eqnarray}
H_{\rm MF} &=& 
{ \sum_{{\bf k} \sigma} }^{\prime}
\left( \xi_k c_{{\bf k}\sigma}^{\dagger} c_{{\bf k}\sigma}
- \xi_k c_{-{\bf k}\sigma} c_{-{\bf k}\sigma}^{\dagger} \right) 
\nonumber \\
& & 
+ {\sum_{ {\bf k} \sigma_1 \sigma_2 }}^{\prime}
\left[ 
\left( \Delta^{\bf k} \right)^{\dagger}_{\sigma_1 \sigma_2}
c_{-{\bf k}\sigma_2} c_{{\bf k}\sigma_1}
+
\Delta^{\bf k}_{\sigma_1 \sigma_2}
c_{{\bf k}\sigma_1}^{\dagger} 
c_{-{\bf k}\sigma_2}^{\dagger} 
\right]. \nonumber \\
\end{eqnarray}
Here the summation in ${\bf k}$ space is taken over a half of the
first Brillouin zone.

For the spin singlet pairing case in which $\Delta^{\bf k}_{\uparrow
\downarrow} = - \Delta^{\bf k}_{\downarrow \uparrow} \equiv
\Delta_{\bf k}$, the gap equation can be derived by taking the
standard procedure:\cite{SIGRIST_UEDA}
\begin{equation}
\Delta_{\bf k} = -\frac{1}{2\Omega} \sum_{{\bf k}'(\neq {\bf k})}
V_{{\bf k}{\bf k}'} \frac{\Delta_{{\bf k}'}}{E_{{\bf k}'}}
\tanh \frac{\beta E_{{\bf k}'}}{2},
\label{gap_eq}
\end{equation}
with $E_{\bf k} = \sqrt{\xi_k^2 + |\Delta_{\bf k}|^2}$.
For the spin triplet pairing case with 
$\Delta_{\uparrow \uparrow}^{\bf k}
= \Delta_{\downarrow \downarrow}^{\bf k} = 0$ and 
$\Delta_{\uparrow \downarrow}^{\bf k}
= \Delta_{\downarrow \uparrow}^{\bf k}$, we obtain the same gap
equation (\ref{gap_eq}). A pairing state with 
$\Delta_{\uparrow \uparrow}^{\bf k} \neq 0$ and/or 
$\Delta_{\downarrow \downarrow}^{\bf k}\neq 0$ may be stabilized in
the presence of an in-plane magnetic field or at the sample's
boundary.
Here we do not consider such a case.

The gap equation (\ref{gap_eq}) is the same as the gap equation for
the composite fermion pairing state at half-filled Landau
levels.\cite{GWW,MORINARI}
We apply the same analysis of Ref.~\cite{GWW}.
We assume that the gap function $\Delta_{\bf k}$ has the following
form:\cite{GWW}
\begin{equation}
\Delta_{\bf k} = \Delta_k \exp \left( -i \ell \theta_{\bf k} \right).
\label{eq_ansatz}
\end{equation}
After substituting Eq.~(\ref{eq_ansatz}) into Eq.~(\ref{gap_eq}), we
integrate the angular variable $\theta_{\bf k}$ using Cauchy's theorem 
by taking $\exp (i\ell \theta_{\bf k})$ as a complex variable.
In this calculation, we find that the attractive interaction arises only
in the case of $\ell > 0$.
From the fact that the case of $\ell < 0$ yields a repulsive
interaction, we may exclude the possibility of a linear combination of
components with $\ell > 0$ and $\ell < 0$ in Eq.~(\ref{eq_ansatz}).

For the ground state, we obtain \cite{GWW}
\begin{equation}
\Delta_k = \frac{1}{2m\theta} \left[
\int_0^k dk' \frac{k'\Delta_{k'}}{E_{k'}}
\left(\frac{k'}{k} \right)^{\ell}
+ \int_k^{\infty} dk' \frac{k'\Delta_{k'}}{E_{k'}} 
\left(\frac{k}{k'} \right)^{\ell}
\right].
\label{eq_gapint}
\end{equation}
In order to solve this non-linear integral equation, we use an
approximation.
From the asymptotic behavior of the right hand side of
Eq.~(\ref{eq_gapint}), we set \cite{GWW}
\begin{equation}
\Delta_k = \left\{
\begin{array}{cc}
\Delta \epsilon_F (k/k_F)^{\ell}, & \hspace{1em} {\rm for}~~k<k_F,\\
\Delta \epsilon_F (k_F/k)^{\ell}, & \hspace{1em} {\rm for}~~k>k_F.
\end{array}
\right.
\end{equation}
Substituting this equation into Eq.~(\ref{eq_gapint}), we
obtain\cite{GWW}
\begin{eqnarray}
\lefteqn{\int_0^{k_F} (dk/k_F)
\frac{\left(k/k_F\right)^{2\ell +1}}{\sqrt{\left[ (k/k_F)^2-1 \right]^2 +
\Delta^2 (k/k_F)^{2\ell}}}} \nonumber \\
& & + 
\int_{k_F}^{\infty} (dk/k_F)
\frac{\left(k/k_F\right)^{1-2\ell}}{\sqrt{\left[ (k/k_F)^2-1 \right]^2 +
\Delta^2 (k/k_F)^{-2\ell}}} = \theta.
\label{eq_gapint2}
\end{eqnarray}
The remaining parameter $\Delta$ can be evaluated numerically 
from Eq.~(\ref{eq_gapint2}).\cite{MORINARI}
For each $\ell$ we estimate the left hand side of
Eq.~(\ref{eq_gapint2}) with varying the value of $\Delta$.
The point at which Eq.~(\ref{eq_gapint2}) is satisfied gives the value 
of $\Delta$.
From this analysis we find that the largest gap is obtained for the
case of $\ell = 1$ and $\Delta^{\ell=1} \sim 3.1$.
Furthermore, this state has the lowest ground state enegy.
Therefore, the ground state is p-wave superconductivity.
From Eq.~(\ref{eq_ansatz}), the symmetry of the Cooper pair is $p_x
\pm i p_y$.
Incidentally, this is the same pairing state as that proposed in
$Sr_2RuO_4$.\cite{RU_p}
However, we cannot apply our pairing mechanism to this system.
We shall discuss this point in Sec.~\ref{sec_discussion}

\section{Antiferromagnetic case}
\label{sec_antiferro}
Now we study the mechanism of the case of
antiferromagnetic coupling between the localized spins.
Although we can apply a similar analysis, the antiferromagnetic case
is more complicated than the ferromagnetic case
because we need to cope with staggered magnetization.
That is, we need to distinguish the A and B sublattices.
Fortunately, there is a transformation by which the system is mapped
onto a similar model of Eq.~(\ref{eq_actionF}) in
Sec.~\ref{sec_ferro}.
We introduce such a transformation and analyze the mechanism of
superconductivity through the transformation.

As in the ferromagnetic case, we introduce Schwinger bosons and rotate 
the carrier's spins so that each of them is in the direction of
the localized spin at the same site.
The rotations are performed by the following transformations:
\begin{equation}
\left\{
\begin{array}{c}
c_i \rightarrow U_i c_i \hspace{2em}(i\in A) \\
c_j \rightarrow U_j \left( -i \sigma_2 \right) c_j \hspace{2em}(j\in B)
\end{array}
\right.
\label{eq_AFU}
\end{equation}
The definition of $U_{\ell}$ is given by Eq.~(\ref{eq_defU}).
Note that the matrix,
\begin{equation}
U_j \left( -i \sigma_2 \right) = \left(
\begin{array}{cc}
-\overline{z}_{j\downarrow} & - z_{j\uparrow} \\
\overline{z}_{j\uparrow} & -z_{j\downarrow}
\end{array}
\right),
\label{eq_UB}
\end{equation}
can be derived from the matrix $U_j$ by the transformation
\begin{equation}
\left\{
\begin{array}{ccc}
z_{j\uparrow} & \rightarrow & -\overline{z}_{j\downarrow} \\
z_{j\downarrow} & \rightarrow & \overline{z}_{j\uparrow}.
\end{array}
\right.
\end{equation}
If we do the same transformation in ${\bf S}_j = \frac12 \overline{z}_j 
{\mbox{\boldmath ${\bf \sigma}$}} z$, we obtain
${\bf S}_j \rightarrow - \frac12 \overline{z}_j 
{\mbox{\boldmath ${\bf \sigma}$}} z$.
Thus, in Eq.~(\ref{eq_AFU})
the presence of the factor $(-i \sigma_2)$ at the B sublattice implies
that the carrier's spin is in the direction of the staggered component
of the localized spins.

In order to eliminate the factor $(-i\sigma_2)$, we perform the
following transformation at the B sublattice:
\begin{equation}
c_j = i\sigma_2 \tilde{c}_j \hspace{2em} (j\in B).
\label{af_f}
\end{equation}
By this transformation, we obtain
\begin{equation}
S_c + S_{\rm int} = \int_0^{\beta} d\tau \int d^2 {\bf r}
\overline{\tilde{\psi}} ({\bf r},\tau) 
G^{-1} (\{ {\hat k}_{\mu} + {\cal A}_{\mu} \})
\tilde{\psi} ({\bf r},\tau),
\end{equation}
where $G^{-1}(\{ k_{\mu} \})$ is given by Eq.~(\ref{eq_Ginv}).
This action has the same form as the action of the ferromagnetic case, 
Eq.~(\ref{eq_cint}).
Therefore, the same Chern-Simons term is induced by integrating
out the carrier fields.
However, we need to perform the inverse transformation of
Eq.~(\ref{af_f}) when we study the symmetry of the Cooper pair because 
the transformation (\ref{af_f}) affects the order parameter of the
Cooper pair.
Furthermore, the action of the localized spin system, of course,
differs from the action of the ferromagnetic case.
In particular, the relevant gauge field component is different from
that case.

\subsection{Action of the localized spin system}
In order to identify which component of ${\cal A}_{\mu}^a$ is
connected with the antiferromagnetic spin fluctuations, we calculate
the action of the localized spin system.
In terms of the Schwinger bosons, the Hamiltonian is written as, up to 
constant,
\begin{equation}
H_{\rm spin} 
= \frac{J}{2}
\sum_{i\in {\rm A}} 
\sum_{{\mbox{\boldmath ${\bf \eta}$}}=(\pm a,0),(0,\pm a)}
\sum_{\sigma_1, \sigma_2}
z_{i\sigma_1}^{\dagger}
z_{i\sigma_2}
z_{i+{\mbox{\boldmath ${\bf \eta}$}},\sigma_2}^{\dagger}
z_{i+{\mbox{\boldmath ${\bf \eta}$}},\sigma_1}.
\end{equation}
To write this Hamiltonian in a tractable way, we perform the following 
transformation at the B sublattice \cite{AROVAS_AUERBACH}
\begin{equation}
\left\{
\begin{array}{ccc}
z_{j\uparrow} & \rightarrow & -z_{j\downarrow}, \\
z_{j\downarrow} & \rightarrow & z_{j\uparrow}.
\end{array}
\right.
\label{eq_AFtr}
\end{equation}
Turning to the path-integral formalism, we obtain
${\cal Z}_{\rm spin} 
= \int {\cal D} \overline{z} {\cal D} z {\cal D} \lambda^{\rm SB}
\exp \left( -S_{\rm spin} \right)$, where
\begin{eqnarray}
S_{\rm spin} &=& \int_0^{\beta} d\tau
\left[
\sum_{\ell} \sum_{\sigma}
\overline{z}_{\ell \sigma} \left( \partial_{\tau} +
\lambda_{\ell}^{\rm SB} \right) z_{\ell \sigma} \right. \nonumber \\
& & \left. -\frac{J}{2} \sum_{i\in {\rm A}} 
\sum_{\mbox{\boldmath ${\bf \eta}$}} 
\sum_{\sigma_1 \sigma_2}
\overline{z}_{i + {\mbox{\boldmath ${\bf \eta}$}},\sigma_1}
\overline{z}_{i\sigma_1}
z_{i\sigma_2}
z_{i + {\mbox{\boldmath ${\bf \eta}$}},\sigma_2} \right],
\end{eqnarray}
where the $\tau$ dependence of all fields is implicit.
In order to decouple the interaction term, we introduce
Stratonovich-Hubbard fields 
$Q_{i,i+{\mbox{\boldmath ${\bf \eta}$}}}$ and 
$\overline{Q}_{i,i+{\mbox{\boldmath ${\bf \eta}$}}}$:
${\cal Z}_{\rm spin} 
= \int {\cal D} \overline{z} {\cal D} z {\cal D} \lambda^{\rm SB}
{\cal D} \overline{Q} {\cal D} Q
\exp \left( -S_{\rm spin} \right)$,
\begin{eqnarray}
S_{\rm spin} &=& \int_0^{\beta} d\tau
\left[
\sum_{\ell} \sum_{\sigma}
\overline{z}_{\ell \sigma} \left( \partial_{\tau} +
\lambda_{\ell}^{\rm SB} \right) z_{\ell \sigma} \right. \nonumber \\
& & \left. 
+\frac{J}{2} \sum_{i\in {\rm A}} 
\sum_{\mbox{\boldmath ${\bf \eta}$}} 
\overline{Q}_{i,i+{\mbox{\boldmath ${\bf \eta}$}}}
Q_{i,i+{\mbox{\boldmath ${\bf \eta}$}}} \right. \nonumber \\
& & \left. - \frac{J}{2} 
\sum_{i\in {\rm A}} \sum_{\mbox{\boldmath ${\bf \eta}$}} 
\sum_{\sigma}
\left(
Q_{i,i+{\mbox{\boldmath ${\bf \eta}$}}}
\overline{z}_{i + {\mbox{\boldmath ${\bf \eta}$}},\sigma}
\overline{z}_{i\sigma} \right. \right. \nonumber \\
& & \left. \left. + \overline{Q}_{i,i+{\mbox{\boldmath ${\bf \eta}$}}}
z_{i\sigma}
z_{i + {\mbox{\boldmath ${\bf \eta}$}},\sigma}
\right)
\right].
\label{after_SH}
\end{eqnarray}
The spin fluctuation field 
$Q_{i,i+{\mbox{\boldmath ${\bf \eta}$}}}$ consists of the phase
fluctuations and the amplitude fluctuations.
The latter is irrelevant for our analysis as it is in
Sec.~\ref{sec_ferro}.
The phase fluctuations are connected with a gauge invariance of
Schwinger bosons.\cite{READ_SACHDEV}
We include these phase fluctuation degrees of freedom later by
imposing the gauge invariance.

We set $Q_{i,i+{\mbox{\boldmath ${\bf \eta}$}}}
=|Q_{i,i+{\mbox{\boldmath ${\bf \eta}$}}}|=Q={\rm const.}$ and
$\lambda_{\ell}^{\rm SB} = \lambda_{\rm SB}={\rm const.}$.
Then, the action is diagonalized in ${\bf k}$-space.
Introducing the following fields,
\begin{equation}
{\zeta}_{{\bf k}\sigma}=\frac12 \left[
\left( z_{{\bf k}\sigma} + z_{{\bf k}+{\bf Q},\sigma} \right)
+ 
\left( \overline{z}_{-{\bf k}\sigma} - \overline{z}_{-{\bf k}+{\bf
Q},\sigma} \right)
\right],
\end{equation}
\begin{equation}
{\Xi}_{{\bf k}\sigma} =\frac12 \left[
\left(
z_{{\bf k}\sigma} + z_{{\bf k}+{\bf Q},\sigma} \right)
-
\left(
\overline{z}_{-{\bf k}\sigma} - \overline{z}_{-{\bf k}+{\bf Q},\sigma}
\right)
\right],
\end{equation}
with ${\bf Q}=(\pi/a,\pi/a)$,
the action is written in terms of these fields as
\begin{eqnarray}
S_{\rm spin}
&=& \sum_{i\omega_n} { \sum_{\bf k} }^{\prime} \sum_{\sigma}
\left[
-i\omega_n 
\left( \overline{\Xi}_{{\bf k}\sigma} \zeta_{{\bf k}\sigma}
+
\overline{\zeta}_{{\bf k}\sigma} \Xi_{{\bf k}\sigma} \right)
\right. \nonumber \\
& & \left. + (\lambda_{\rm SB}+\epsilon_{\bf k})
\overline{\zeta}_{{\bf k}\sigma} \zeta_{{\bf k}\sigma}
+ (\lambda_{\rm SB}-\epsilon_{\bf k})
\overline{\Xi}_{{\bf k}\sigma} {\Xi}_{{\bf k}\sigma}
\right],
\end{eqnarray}
where $\epsilon_{\bf k}=-2JQ [ \cos (k_x a) + \cos (k_y a) ]$ and the
summation in ${\bf k}$ space is taken over a half of the first
Brillouin zone.
One can see that the mass of $\Xi_{{\bf k} \sigma}$ is 
$\lambda_{\rm SB} + 2JQ$, which is larger than the mass of 
$\zeta_{{\bf k}\sigma}$,
$\lambda_{\rm SB}-2JQ$. \cite{READ_SACHDEV}
Furthermore, the mass of $\Xi_{{\bf k} \sigma}$ is non-vanishing
whereas the mass of $\zeta_{{\bf k}\sigma}$ is identically zero in
the ordered phase.
Therefore, we can safely integrate out 
$\Xi_{{\bf k} \sigma}$ and we obtain
\begin{equation}
S_{\rm spin}
= \sum_k \sum_{\sigma}
\left[ - \frac{(i\omega_n)^2}{\lambda_{\rm SB}-\epsilon_{\bf k}}
+ \lambda_{\rm SB} + \epsilon_{\bf k} \right]
\overline{\zeta}_{k\sigma} 
\zeta_{k\sigma}.
\end{equation}

Taking the continuum limit and recovering the gauge invariance of the 
Schwinger bosons, we obtain
\begin{eqnarray}
S_{\rm spin} &=& \frac{2}{g}
\int_0^{\beta c_{\rm sw}} \!\!\!\! dx_0 \int d^2 {\bf x} 
\sum_{\sigma}
\left[ |\left( \partial_{\mu} + i {\cal A}_{\mu}^{\rm SB} \right) 
\zeta_{\sigma}|^2 + \frac{\Delta_{\rm sw}^2}{c^2_{\rm sw}}
|\zeta_{\sigma}|^2 \right],\nonumber \\
\label{eq_spinpre}
\end{eqnarray}
where
$g=4\sqrt{2} a$,
$\Delta_{\rm sw} = \sqrt{\lambda_{\rm SB}^2-4J^2 Q^2}$,
$c_{\rm sw}=\sqrt{2} JQ a$,
and $x_0 = c_{\rm sw} \tau$.
In these parameters we set $\lambda_{\rm SB} = 2JQ$ except for 
$\Delta_{\rm sw}$.
Equation (\ref{eq_spinpre}) is invariant under the gauge
transformation\cite{READ_SACHDEV}
\begin{equation}
\zeta ({\bf r},\tau ) \rightarrow \zeta ({\bf r},\tau ) 
\exp \left[ i\theta ({\bf r},\tau ) \right],
\end{equation}
\begin{equation}
{\cal A}_{\mu}^{\rm SB} ({\bf r},\tau)
\rightarrow
{\cal A}_{\mu}^{\rm SB} ({\bf r},\tau)
- \partial_{\mu} \theta ({\bf r},\tau).
\end{equation}
This gauge transformation corresponds to 
\begin{equation}
\left\{
\begin{array}{cccc}
z_i &\rightarrow &z_i \exp (i\theta_i) & (i\in A),\\
z_j &\rightarrow &z_j \exp (-i\theta_j) & (j\in B),
\end{array}
\right.
\end{equation}
because if we take the set of $({\rm even}, {\rm even})$ and 
$({\rm odd},{\rm odd})$ for the A sublattice, then
\begin{equation}
\left\{
\begin{array}{c}
{\zeta}_{{\ell}\sigma}=z_{{\ell}\sigma} \hspace{1em}
{\rm for}~~\ell \in {\rm A}, \\
{\zeta}_{{\ell}\sigma}=\overline{z}_{{\ell}\sigma}
\hspace{1em}{\rm for}~~\ell \in {\rm B}.
\end{array}
\right.
\end{equation}
This equation is verified as follows
\begin{eqnarray}
{\zeta}_{\ell \sigma}
&=& \frac{1}{N} {\sum_{\bf k}}^{\prime} {\zeta}_{{\bf k}\sigma}
\exp (i{\bf k}\cdot {\bf R}_{\ell} ) \nonumber \\
&=& 
\frac{1}{2N} {\sum_{\bf k}}^{\prime}
z_{{\bf k}\sigma} {\rm e}^{i{\bf k}\cdot {\bf R}_{\ell}}
\left( 1 + {\rm e}^{-i {\bf Q} \cdot {\bf R}_{\ell}} \right)
\nonumber \\
& & +
\frac{1}{2N} {\sum_{\bf k}}^{\prime}
\overline{z}_{-{\bf k}\sigma} {\rm e}^{i{\bf k}\cdot {\bf R}_{\ell}}
\left( 1 - {\rm e}^{-i {\bf Q} \cdot {\bf R}_{\ell}} \right).
\end{eqnarray}

Since the gauge field is connected with the phase fluctuations of 
$Q_{i,i+{\mbox{\boldmath ${\bf \eta}$}}}$, the gauge field ${\cal
A}_{\mu}^{\rm SB}$ has the following form
\begin{equation}
{\cal A}_{\mu}^{\rm SB} = -i \sum_{\sigma} \overline{\zeta}_{\sigma} (x)
\partial_{\mu} \zeta_{\sigma} (x).
\end{equation}
In order to find the relationship between ${\cal A}_{\mu}^{\rm SB}$
and ${\cal A}_{\mu}$, we write ${\cal A}_{\mu}$ in terms of
$\zeta_{\ell \sigma}$ and $\overline{\zeta}_{\ell \sigma}$.
Thus, we find 
${\mbox{\boldmath ${\bf {\cal A}}$}}^1 = - {\mbox{\boldmath ${\bf
{\cal A}}$}}^{\rm SB}$.
From the gauge invariance of the Schwinger bosons, one can see that
there is a correspondence between 
${\cal A}_{\tau}^1$ and ${\cal A}_{\tau}^{\rm SB}$.
Therefore, ${\cal A}_{\mu}^1$ is connected with the antiferromagnetic
spin fluctuations.

As a result, we may write the effective action in the following form
\begin{eqnarray}
S &=& S_c + S_{\rm spin} + S_{\rm CS} \nonumber \\
&=& \int_0^{\beta} d\tau \int d^2 {\bf r}
\overline{\tilde{\psi}} ({\bf r},\tau) \nonumber \\
& & \left[
\partial_{\tau} + i{\cal A}_{\tau}^1-\mu 
+\frac{1}{2m} \left( -i \nabla + 
{\mbox{\boldmath ${\bf {\cal A}}$}}^1 \right)^2
\right]
\tilde{\psi} ({\bf r},\tau) \nonumber \\ 
& & + \frac{2}{g} \int d^3 x \sum_{\sigma}
\left[
|\left( \partial_{\mu} + i {\cal A}_{\mu}^1 \right)
\zeta_{\sigma}(x)|^2
+ \frac{\Delta_{\rm sw}^2}{c_{\rm sw}^2} |\zeta_{\sigma}(x)|^2
\right] \nonumber \\
& & - \frac{i\theta}{2\pi} \int_0^{\beta} d\tau \int d^2 {\bf r}
{\cal A}_{\tau}^1 \left( \partial_x A_y^1-\partial_y A_x^1 \right).
\end{eqnarray}

\subsection{Skyrmion excitations}
\label{sec_skyAF}
As discussed in Sec.~\ref{sec_skyF}, carriers are connected with
skyrmion excitations in the localized spin system through the
Chern-Simons term.
However, the connection is slightly different from the
ferromagnetic case because the spin fluctuations are described by the
gauge field ${\cal A}_{\mu}^1$ instead of ${\cal A}_{\mu}^3$.
If we take the $1$-axis for the quantization axis in spin space, the
relationship between the carrier and the skyrmion excitation is 
\begin{equation}
\sum_{\sigma} s_{\sigma} 
\overline{{\tilde{\psi}}}_{\sigma} ({\bf r},\tau)
{\tilde{\psi}}_{\sigma} ({\bf r},\tau)
= -\frac{\theta}{2\pi} \left(
\partial_x {\cal A}_y^1 - \partial_y {\cal A}_x^1 \right).
\label{eq_skAF}
\end{equation}

Contrary to the ferromagnetic case, a significant feature appears for 
the antiferromagnetic case,
that is, a pinning mechanism of carriers in the antiferromagnetic
long-range ordered phase.
This can be seen as follows.
The relationship (\ref{eq_skAF}) is obtained after the
transformation (\ref{af_f}).
In order to capture the proper nature of skyrmion excitations, we must 
go back to the frame before the transformation (\ref{af_f}).
Performing the inverse transformation of Eq.~(\ref{af_f}) at the B
sublattice, we find that an additional sign change is brought about
in the left hand side of Eq.~(\ref{eq_skAF}),
that is, a skyrmion (anti-skyrmion) excitation transformed into an
antiskyrmion (skyrmion) excitation.
Therefore skyrmions or antiskyrmions cannot move to the nearest
neighbor sites as long as there is antiferromagnetic long-range order
and skyrmion excitations have a gap.
This suggests an insulating behavior of the carriers in the
antiferromagnetic long-range ordered phase.

Although this is a new feature which appears in the antiferromagnetic
case, the destruction mechanism of magnetic long-range order is the
same as that discussed in Sec.~\ref{sec_skyF}.
Antiferromagnetic long-range order is destroyed by carrier doping
because carriers behave like skyrmion excitations.
After antiferromagnetic long-range order is destroyed, skyrmion
excitations become gapless excitations.
The Chern-Simons term plays a dominant role and the attractive
interaction between skyrmions leads to superconductivity.

\subsection{Superconductivity}
\label{sec_afsc}
Now we investigate the property of superconductivity.
In order to identify the symmetry of the Cooper pair, we need to
perform the inverse transformation of Eq.~(\ref{af_f}).
However, since the calculation of the pairing matrix is much easier
for the system after the transformation (\ref{af_f}),
we investigate the pairing matrix through the transformation
(\ref{af_f}).
If the B sublattice is consists of the sites with 
$({\rm even}, {\rm odd})$ or $({\rm odd},{\rm even})$, then
the transformation (\ref{af_f}) is written in ${\bf k}$-space as
\begin{equation}
c_{\bf k}-c_{{\bf k}+{\bf Q}}
=
i\sigma_2 \left( \tilde{c}_{\bf k}-\tilde{c}_{{\bf k}+{\bf Q}}
\right).
\label{af_f2}
\end{equation}
Since there is no change at the A sublattice, that is, 
$c_{\bf k}+c_{{\bf k}+{\bf Q}}=
\tilde{c}_{\bf k}+\tilde{c}_{{\bf k}+{\bf Q}}$, we obtain
\begin{eqnarray}
c_{\bf k} &=& \frac12 ~\left[ (\sigma_0+i\sigma_2)\tilde{c}_{\bf k}
+ (\sigma_0 - i\sigma_2)\tilde{c}_{{\bf k}+{\bf Q}} \right], \\
c_{{\bf k}+{\bf Q}} &=& \frac12 ~\left[ (\sigma_0-i\sigma_2)\tilde{c}_{\bf k}
+ (\sigma_0 + i\sigma_2)\tilde{c}_{{\bf k}+{\bf Q}} \right].
\end{eqnarray}

In order to apply the analysis done in Sec.~\ref{sec_sc}, we need to
introduce the following fields
\begin{equation}
\left(
\begin{array}{c}
\tilde{c}_{{\bf k}\uparrow} \\
\tilde{c}_{{\bf k}\downarrow}
\end{array} \right)
= \frac{1}{\sqrt{2}}
\left( \begin{array}{cc}
1 & 1 \\
1 & -1 
\end{array}
\right)
\left(
\begin{array}{c}
\tilde{\chi}_{{\bf k}\uparrow} \\
\tilde{\chi}_{{\bf k}\downarrow}
\end{array} \right),
\end{equation}
because in Eq.~(\ref{eq_skAF})
we take the $1$-axis for the quantization axis in spin space.
In terms of these fields $\tilde{\chi}_{{\bf k}\sigma}$, 
the spin singlet order parameter is written as
\begin{eqnarray}
\lefteqn{
\langle 
c_{-{\bf k}\downarrow} c_{{\bf k}\uparrow}
-
c_{-{\bf k}\uparrow} c_{{\bf k}\downarrow} 
\rangle} \nonumber \\
&=&
\frac12
\left[
\langle 
\tilde{\chi}_{-{\bf k}\downarrow} \tilde{\chi}_{{\bf k}\uparrow}
-
\tilde{\chi}_{-{\bf k}\uparrow} \tilde{\chi}_{{\bf k}\downarrow} 
\rangle \right. \nonumber \\
& & \left. + \langle 
\tilde{\chi}_{-{\bf k}+{\bf Q} \downarrow} \tilde{\chi}_{{\bf k}+{\bf Q}
\uparrow}
-
\tilde{\chi}_{-{\bf k}+{\bf Q} \uparrow} \tilde{\chi}_{{\bf k}+{\bf Q}
\downarrow} 
\rangle
\right].
\end{eqnarray}

For the spin singlet pairing state, the gap function $\Delta_{\bf k}$
is given by
\begin{equation}
\Delta_{\bf k} = \left(\tilde{\Delta}_{\bf k}^{(1)} + 
\tilde{\Delta}_{\bf k}^{(1)} \right)/2,
\end{equation}
where $\tilde{\Delta}_{\bf k}^{(1)}$ is the gap function for 
$\langle \tilde{\chi}_{-{\bf k}\downarrow} 
\tilde{\chi}_{{\bf k}\uparrow} \rangle
= - \langle \tilde{\chi}_{-{\bf k}\uparrow} 
\tilde{\chi}_{{\bf k}\downarrow} \rangle$ and
$\tilde{\Delta}_{\bf k}^{(2)}$ is that for
$\langle 
\tilde{\chi}_{-{\bf k}+{\bf Q} \downarrow} 
\tilde{\chi}_{{\bf k}+{\bf Q}
\uparrow} \rangle = 
- \langle 
\tilde{\chi}_{-{\bf k}+{\bf Q} \uparrow} 
\tilde{\chi}_{{\bf k}+{\bf Q}
\downarrow} \rangle$.
In the continuum limit, 
$\tilde{\Delta}_{\bf k}^{(1)}$ satisfies, at $T=0$,
\begin{equation}
\tilde{\Delta}_{\bf k}^{(1)}
= \frac{1}{2\Omega} \sum_{{\bf k}'(\neq {\bf k})}
V_{{\bf k}{\bf k}'}
\frac{\tilde{\Delta}_{\bf k}^{(1)}}
{E_{{\bf k}'}^{(1)}},
\label{eq_gap}
\end{equation}
where $E_{\bf k}^{(1)} = 
\sqrt{\xi_k^2+|\tilde{\Delta}_{\bf k}^{(1)}|^2}$.
On the other hand, $\tilde{\Delta}_{{\bf k}}^{(2)}$ 
satisfies
\begin{equation}
\tilde{\Delta}_{\bf k}^{(2)}
= - \frac{1}{2\Omega} \sum_{{\bf k}'(\neq {\bf k})}
V_{{\bf k}{\bf k}'}
\frac{\tilde{\Delta}_{\bf k}^{(2)}}
{E_{{\bf k}'}^{(2)}},
\label{eq_gap2}
\end{equation}
where $E_{\bf k}^{(2)} = 
\sqrt{\xi_k^2+|\tilde{\Delta}_{\bf k}^{(2)}|^2}$.
The minus sign in the right hand side of Eq.~(\ref{eq_gap2}) comes
from the sign change in minimal coupling to the gauge field.
The minimal coupling term is derived from 
$\left\{ \partial_{{\cal A}_{\mu}} G_{\rm K.E.}^{-1}
(\{ \hat{k}_{\mu}+{\cal A}_{\mu} \})|_{{\cal A}_{\mu}=0}, 
{\cal A}_{\mu} \right\}$,
where $G_{\rm K.E.}^{-1} (\{ k_{\mu} \})$ represents the kinetic
energy part of the inverse of Green's function 
$G^{-1} (\{ k_{\mu} \})$.
In the lattice system, the displacement of ${\bf Q}$ in ${\bf k}$
space yields the sign change because,
$-2t \left[ \cos \left( k_x + Q_x \right)a
+ \cos \left( k_y + Q_y \right)a \right]
= 
+ 2t \left( \cos k_xa + \cos k_ya \right)$.
This sign change affects the interaction term through the minimal
coupling term.
Equation (\ref{eq_gap2}) is understood to be the equation obtained
after taking the continuum limit.

The analysis of Eqs. (\ref{eq_gap}) and (\ref{eq_gap2}) parallels that 
in Sec.~\ref{sec_sc}.
So we also take Eq.~(\ref{eq_ansatz}) for the form of 
$\tilde{\Delta}_{\bf k}^{(1)}$ and $\tilde{\Delta}_{\bf k}^{(2)}$.
From Eqs. (\ref{eq_gap}) and (\ref{eq_gap2}), we see that 
$\tilde{\Delta}_{\bf k}^{(1)}$ and 
$\tilde{\Delta}_{\bf k}^{(2)}$ have opposite chirality.
For the case of the spin singlet pairing, $|\ell|$ takes the values
$|\ell |=2,4,6,\cdots$.
From the same analysis done in Sec.~\ref{sec_sc}, we find that the
pairing state with the lowest energy is $|\ell|=2$, or d-wave.
The pairing function is given by
\begin{eqnarray}
\Delta_{\bf k}
&=& \tilde{\Delta}_k 
\left[ 
\exp \left( 2i \theta_{\bf k} \right)
+ \exp \left( -2i \theta_{\bf k} \right) \right]/2 \nonumber \\
&=& \tilde{\Delta}_k \left( \cos^2 \theta_{\bf k}
- \sin^2 \theta_{\bf k} \right),
\label{eq_dwave}
\end{eqnarray}
where $\tilde{\Delta}_k = \tilde{\Delta} \tilde{\epsilon}_F
(k_</k_>)^2$
with $\tilde{\Delta} \sim 1.3$.
Here if $k$ is smaller than $k_F$ then $k_< = k$ and $k_> =k_F$ and
vice versa.
From Eq.~(\ref{eq_dwave}), we see that the precise symmetry of the
Cooper pair is $d_{x^2-y^2}$.

For the spin triplet pairing case, we find that 
$\Delta^{\bf k}_{\uparrow \uparrow}
=\Delta^{\bf k}_{\downarrow \downarrow} = 0$ and
$\Delta^{\bf k}_{\uparrow \downarrow}
=\Delta^{\bf k}_{\downarrow \uparrow} 
\propto \sin \theta_{\bf k}$.
In the d-vector notation of the triplet pairing state, \cite{LEGGETT}
this is written as ${\bf d}_{\bf k} = k_y \hat{e}_3$.
From the analysis of the Ginzburg-Landau free energy \cite{RU_p}
within the two-dimensional representation in which the basis states
are $\{k_x \hat{e}_3, k_y \hat{e}_3\}$, one can see that 
the free energy of this pairing state is higher than that of 
${\bf d}_{\bf k}=(k_x \pm i k_y) \hat{e}_3$.
This suggests that the pairing state with ${\bf d}_{\bf k} = k_y
\hat{e}_3$ is not stabilized and we can exclude the possiblity of the
spin triplet pairing state.

From the above analysis, we may conclude that the symmetry of the
Cooper pair is d-wave, or more precisely $d_{x^2-y^2}$-wave.
This is the same pairing state as that of the high-$T_c$
superconductors.

\section{Discussion}
\label{sec_discussion}
In this section, we discuss the condition for that the Chern-Simons
term is induced and applications to the high-$T_c$ cuprates and other
systems.

\subsection{Condition for the Chern-Simons term}
\label{sec_condition}
The derivation of the Chern-Simons term is based on the continuum
approximation. 
This approximation is justified when the length scale of
the gauge field is relatively larger than the length scale determined
by spin-orbit coupling.
We first discuss the validity of this assumption.

As we have shown in Sec.~\ref{sec_so}, the effect of spin-orbit
coupling is to rotate the carrier's spin at every hopping process.
The angle of this rotation is $2\lambda/t (\equiv \theta_{\rm so})$
for the nearest-neighbor hopping.
On the other hand, the length scale of the gauge field is determined
by the fluctuations of the spin system through Eq.~(\ref{eq_A}).
This length scale is given by $\hbar c_{\rm sw}/(\pi \Delta_{\rm
sw})$ from the analysis of the Schwinger bosons and
it is translated into the fluctuation angle of the localized spins.
For the nearest neighbor sites, the fluctuation angle is,
$\sim a/(\hbar c_{\rm sw}/\pi \Delta_{\rm sw}) = \pi \Delta_{\rm
sw}/(\hbar c_{\rm sw} /a)$.
In order to apply the continuum approximation to the gauge field, this 
fluctuation angle should be smaller than $\theta_{\rm so}=2\lambda/t$,
that is,
\begin{equation}
\frac{2\lambda}{t} \gg \frac{\pi \Delta_{\rm sw}}{(\hbar c_{\rm sw}/a)}.
\label{eq_cond}
\end{equation}
This is the condition of taking the continuum limit for the gauge
field.

The condition (\ref{eq_cond}) is satisfied as long as sufficient
magnetic correlations are preserved, or the gap of spin wave
excitations is relatively small.
Note that the condition (\ref{eq_cond}) is just for the localized spin
system. 
It only provides the condition for the existence of the Chern-Simons 
term.
Although our mechanism of superconductivity is based on the presence
of the Chern-Simons term, the superconducting state 
does not rely on directly the value of spin-orbit coupling $\lambda$.
In fact, the gap of superconductivity is independent of the parameter
$\lambda$ as shown in Secs.~\ref{sec_sc} and ~\ref{sec_afsc}.

\subsection{High-$T_c$ cuprates}
As we have discussed in Introduction, the high-$T_c$ cuprates can be
described as the Kondo lattice system. 
In addition, buckling seems to play an
important role concerning the occurrence of superconductivity.
If buckling and the parameters characterizing the localized spin
system fulfill the condition (\ref{eq_cond}),
then the Chern-Simons term is induced.

If we can apply our mechanism to the high-$T_c$ cuprates, then it
describes the underdoped region.
Because the underdoped region is close to the antiferromagnetic
long-range ordered phase.
Such a phase is properly described by the Schwinger bosons.
However, in the optimal doped region and the overdoped region of the
high-$T_c$ cuprates the localized spin system is much more disordered
than in the underdoped region.

In order to describe strongly disordered regions, the Schwinger bosons 
are not appropriate fields.
The description in terms of fermionic fields is more suitable than the 
bosonic description of the localized spin system.
The description by fermionic fields instead of the Schwinger bosons
corresponds to describe the $Cu$ site degrees of freedom by fermion
fields. 
Such fermionic degrees of freedom may be observed by angle-resolved
photoemission spectroscopy experiments (ARPES).

It should be noted that the antiferromagnetic correlation 
described in terms of fermion fields has the form of the singlet
pairing between the fermions,
because the order parameter for the antiferromagnetic correlations, 
$Q_{i,i+{\mbox{\boldmath ${\bf \eta}$}}}$ in the Schwinger boson
system is replaced by $\langle f_{i\uparrow}f_{j\downarrow}
-f_{i\downarrow}f_{j\uparrow} \rangle \equiv f_{ij}$,
in terms of the fermion fields, $f_{i\sigma}$.
In deriving this relationship we need to go back to the original
system before the transformation of Eq.~(\ref{eq_AFtr}) and we use the 
constraint $\sum_{\sigma} f_{i\sigma}^{\dagger} f_{i\sigma}=1$.
This correlation is similar to that 
characterizes the RVB state.\cite{ANDERSON_ETAL}
From the mean field analysis, we find that $f_{ij}$ shows 
$d_{x^2-y^2}$ symmetry.\cite{KOTLIAR_LIU}
Although this $d_{x^2-y^2}$ symmetry is the same as that of the Cooper
pair of holes, they have completely different origins and must be
discussed independently.

Now we discuss some properties of the superconducting state.
In our mechanism, there is a characteristic excitation.
Because of the transformation (\ref{af_f2}), two
components of Cooper pairs appears:
$\langle \tilde{\chi}_{-{\bf k}\downarrow} 
\tilde{\chi}_{{\bf k}\uparrow} \rangle$
and 
$\langle \tilde{\chi}_{-{\bf k}+{\bf Q}\downarrow} 
\tilde{\chi}_{{\bf k}+{\bf Q}\uparrow} \rangle$.
Therefore, there is an excitation between them that
creates one quasi-hole in the one component of the Cooper pairs and
one quasiparticle in the other with the momentum ${\bf
Q}=(\pi/a,\pi/a)$ and the energy $2\tilde{\Delta}_{k_F}$.
Such an excitation can be detected by inelastic neutron scattering.
This excitation may be identified with the 41meV peak observed by the
neutron experiments.\cite{NEUTRON_41}

In addition, it should be stressed that the strength of spin-orbit
coupling does not affect the transition temperature of
superconductivity as long as the condition for the presence of the
Chern-Simons term is satisfied, as discussed in the last subsection.

Buckling affects superconductivity not through the spin-orbit coupling
term but through the Fermi energy because the superconducting gap is
proportional to the Fermi energy.
If we increase the angle of buckling, then the hopping amplitude $t$
in the plane may be reduced.
Such a reduction leads to decrease of the Fermi energy.
Therefore, if we increase the angle of buckling, the transition
temperature is rather reduced.
This is consistent with the experiments.\cite{CHMAISSEM_ETAL}

\subsection{Double-exchange systems}
Our mechanism can be applied to the double-exchange system.
However, in application to that system the following conditions
should be satisfied.
First, we must detect superconductivity in the region where the
antiferromagnetic correlation between the core spins are preserved.
That is, the carrier number must be small.
Whereas the region in which the ferromagnetic correlation between the
core spins dominates, we cannot apply our mechanism.
Secondly, we require the system with the layered structure and
buckling of the planes.
Although $La_{2-2x}Sr_{1+2x}Mn_2O_7$ is a layered
double-exchange system, the $x=0.3$ compound shows ferromagnetic
correlation.\cite{KIMURA_ETAL}
If the compound with buckling and small $x$ in which the sample shows
antiferromagnetic correlation is provided, $d_{x^2-y^2}$
superconductivity based on our mechanism will be realized.

\subsection{Other systems}
From the symmetry of the Cooper pair in Sec.~\ref{sec_sc}, one might
think of the application to $Sr_2RuO_4$.
However, the magnetism relevant to this system is itinerant
ferromagnetism.
In fact, all of the relevant d-orbitals $d_{xy}$, $d_{yz}$, and
$d_{zx}$ form conduction bands and the Fermi surface of them are
observed by the de Haas-van Alphen effect.\cite{RU_FERMI}
In such a system, we cannot expect the formation of the localized
spins.
Therefore, we cannot apply our mechanism to $Sr_2RuO_4$.

For the application of the mechanism of Sec.~\ref{sec_ferro} to real
materials, we require a ferromagnetic superexchange interaction
between the localized spins.
In order to produce such a superexchange interaction, we need at least
three kinds of ions or multi-band structure for the magnetic ions to
constitute the conduction layers. Furthermore, of course, we need
buckling of the planes.

At present, it seems that there is no material with all of these
properties.
However, if such a system exists, then we can expect higher
superconductiving transition temperature than the high-$T_c$ cuprates.
Because the gap is larger than that of the antiferromagnetic case
within the analysis in which the long-range Coulomb interaction is
neglected.

\acknowledgments

This work was supported in part by a Grant-in-Aid from the Ministry of 
Education, Culture, Sports, Science and Technology.

\appendix
\section*{Induced Chern-Simons term}
\label{ap_cs}
In this appendix, we derive the effective action for the gauge field
arising from the carrier system.
Integrating out the carrier fields from the action (\ref{eq_cint}),
we obtain the effective action $S_{\cal A}^c$:
\begin{equation}
S_{\cal A}^c
= - {\rm Tr} \left\{ \ln \left[
G^{-1} ({\hat k}_{\mu} + {\cal A}_{\mu} ) \right]
- \ln \left[ G^{-1} ({\hat k}_{\mu}) \right] \right\}.
\end{equation}
We expand this action with respect to ${\cal A}_{\mu}$ as
$S_{\cal A}^c = S^{(2)} + S^{(3)} + \cdots$, where
\begin{eqnarray}
S^{(2)}
&=& \frac12 {\rm Tr}
\left[
G ~\frac{1}{2} \left\{ 
\frac{\partial G^{-1}}{\partial {\hat k}_{\mu}}, 
{\cal A}_{\mu} \right\} G~ \frac{1}{2} \left\{ 
\frac{\partial G^{-1}}{\partial {\hat k}_{\nu}}, 
{\cal A}_{\nu}\right\} \right], \\
S^{(3)}
&=& - \frac13 {\rm Tr}
\left[
G ~\frac{1}{2} \left\{ 
\frac{\partial G^{-1}}{\partial {\hat k}_{\mu}}, 
{\cal A}_{\mu} \right\} 
G ~\frac{1}{2} \left\{ 
\frac{\partial G^{-1}}{\partial {\hat k}_{\nu}}, 
{\cal A}_{\nu} \right\} \right. \nonumber \\
& & \left. G ~\frac{1}{2} \left\{ 
\frac{\partial G^{-1}}{\partial {\hat k}_{\lambda}}, 
{\cal A}_{\nu} \right\} \right].
\end{eqnarray}

In order to calculate $S^{(2)}$, we apply the derivative expansion
technique.
To illustrate this technique, let us consider a simple one dimensional 
example:
$J={\rm Tr} \left[ V_1 ({\hat k}) F_1 (x) 
V_2 ({\hat k}) F_2 (x) \right]$,
where ${\hat k}=-i\partial_x$ and 
$V_j({\hat k})$ and $F_j(x)$ are functions of ${\hat k}$ and $x$,
respectively.
Inserting the identities 
$\int dx |x\rangle \langle x| = {\hat 1}$ and 
$\sum_k |k\rangle \langle k| = {\hat 1}$, we obtain
\begin{eqnarray}
J & = & \int dx \sum_k 
\langle x| V_1 ({\hat k}) F_1 (x) |k \rangle
V_2(k) F_2(x) \langle k| x \rangle \nonumber \\
&=& \int dx \sum_k F_2(x) V_2(k)
\sum_{n=0}^{\infty}
\frac{V_1^{(n)}(0)}{n!} \langle x| {\hat k}^n F_1(x) |k \rangle
\langle k | x \rangle \nonumber \\
&=& \int dx \sum_k F_2(x) V_2(k) \nonumber \\
& & \times
\sum_{n=0}^{\infty} 
\frac{V_1^{(n)}(0)}{n!} 
\sum_{m=0}^{n} \frac{n!(-i)^m}{m!(n-m)!} 
F_1^{(m)}(x) k^{n-m} \langle x|k \rangle
\langle k | x \rangle \nonumber \\
&=& \frac{1}{L} \sum_k V_1(k)V_2(k) 
\int dx F_1(x) F_2(x) \nonumber \\
& & -\frac{i}{2L} \sum_k 
\left[
V_1^{\prime}(k) V_2(k) 
\int dx F_1^{\prime} (x) F_2 (x) \right. \nonumber \\
& & \left. + V_1(k) V_2^{\prime}(k) 
\int dx F_1 (x) F_2^{\prime} (x) 
\right]
\nonumber \\
& & \quad - \frac{1}{2L} \sum_k V_1^{\prime \prime} (k) V_2 (k)
\int dx F_1^{\prime \prime} (x) F_2 (x) + \cdots.
\end{eqnarray}
In the second term in the last line we have
taken into account the term obtained by the partial integral.
Applying this derivative expansion and retaining the most dominant
term in long-wavelength and low-energy physics,
the action $S^{(2)}$ is evaluated
as
\begin{equation}
S^{(2)} = \frac{i}{4\pi} \sum_{a} \sum_{\mu \nu \lambda}
I_{\mu \nu \lambda}^{aa}
\int_0^{\beta} d\tau \int d^2 {\bf r} {\cal A}_{\mu}^{a}
\partial_{\nu} {\cal A}_{\lambda}^a,
\end{equation}
where 
\begin{equation}
I_{\mu \nu \lambda}^{aa}
= \frac{\pi}{2\Omega} \sum_k {\rm tr}
\left[
G \left\{ \frac{\partial G^{-1}}{\partial k_{\mu}}, \sigma_a \right\}
G \frac{\partial G^{-1}}{\partial k_{\nu}} G
\left\{ \frac{\partial G^{-1}}{\partial k_{\lambda}}, \sigma_a
\right\}
\right].
\end{equation}
Here the trace is taken over spin space.
All functions $I_{\mu \nu \lambda}^{aa}$ can be calculated in a
similar way.
For example, $I_{xy\tau}^{11}$ is given by 
\begin{equation}
I_{xy\tau}^{11}
= 2\pi 
\left( \lambda_1^{(a,0)}\lambda_2^{(0,a)}
- \lambda_2^{(a,0)}\lambda_1^{(0,a)} \right)
J_c K.
\label{eq_2nd}
\end{equation}
In deriving this equation, we have taken into account another choice
of $\lambda^{({\mbox{\boldmath ${\bf \eta}$}})}_a$ connecting
coordinate space with spin space.
In Eq.~(\ref{eq_2nd}), $K$ is given by
\begin{eqnarray}
K &=& \frac{1}{\Omega}\sum_k \frac{1}{\left[
(ik_{\tau}+\xi_k)^2-|\bf g({\bf k})|^2 \right]^2} \nonumber \\
&=& \int \frac{d^2 {\bf k}}{(2\pi)^2}
\left[ 
\frac{
f^{\prime} (\xi_k+|{\bf g}({\bf k})|)
+f^{\prime} (\xi_k-|{\bf g}({\bf k})|)}
{4|{\bf g}({\bf k})|^2} \right. \nonumber \\
& & \left. + \frac{
f(\xi_k-|{\bf g}({\bf k})|)
-f (\xi_k+|{\bf g}({\bf k})|)}
{4|{\bf g}({\bf k})|^3}
\right].
\end{eqnarray}
In the limit of $\beta |J_c| \rightarrow \infty$, this equation is
reduced to 
\begin{equation}
K=\frac{1}{8\pi \lambda |J_c|} \tanh \frac{\beta|J_c|}{8}.
\end{equation}
Thus, $S^{(2)}$ has the form of the Chern-Simons term,
\begin{eqnarray}
S_{CS}^{(2)} 
&=&  -\frac{i\theta}{2\pi} \int_0^{\beta} d\tau \int d^2 {\bf r}
\left[ 
{\cal A}_{\tau}^1 \left( \partial_x {\cal A}_y^1 - 
\partial_y {\cal A}_x^1 \right) \right. \nonumber \\
& & \left. + {\cal A}_{\tau}^2 \left( \partial_x {\cal A}_y^2 - 
\partial_y {\cal A}_x^2 \right)
\right].
\end{eqnarray}
The coefficient of the Chern-Simons term is given by
\begin{equation}
\theta = \frac12 {\rm sgn} \left( J_c \Lambda \right)
\tanh \frac{\beta|J_c|}{8},
\end{equation}
with $\Lambda=\lambda_1^{(a,0)}\lambda_2^{(0,a)}
-\lambda_2^{(a,0)}\lambda_1^{(0,a)}$.

From similar calculations, we find that 
$S^{(3)}$ has a form of a non-Abelian Chern-Simons term,
\begin{equation}
S_{CS}^{(3)} = \frac{i\theta}{\pi}
\int_0^{\beta} d\tau \int d^2 {\bf r}
{\cal A}_{\tau}^3 \left( {\cal A}^1_x {\cal A}^2_y
- {\cal A}^2_x {\cal A}^1_y \right).
\end{equation}
This non-Abelian Chern-Simons term can be reduced to an Abelian
Chern-Simons term upon using the curl free condition,
$2({\cal A}_x^1 {\cal A}_y^2 - {\cal A}_x^2 {\cal A}_y^1)
= \partial_x {\cal A}_y^3 - \partial_y {\cal A}_x^3$:
\begin{equation}
S_{CS}^{(3)} = \frac{i\theta}{2\pi}
\int_0^{\beta} d\tau \int d^2 {\bf r}
{\cal A}_{\tau}^3 \left( \partial_x {\cal A}_y^3 - 
\partial_y {\cal A}_x^3 \right)
\end{equation}

As a result, we may write the effective action for the gauge field
arising form the carrier system in the following form:
\begin{eqnarray}
S_{\cal A}^c
&=& \frac{i\theta}{2\pi} \int_0^{\beta} d\tau \int d^2 {\bf r}
\left[
{\cal A}_{\tau}^3 (\partial_x {\cal A}_y^3 - \partial_y {\cal A}_x^3 )
\right. \nonumber \\ & & \left. 
- {\cal A}_{\tau}^1 (\partial_x {\cal A}_y^1-\partial_y
{\cal A}_x^1)
- {\cal A}_{\tau}^2 (\partial_x {\cal A}_y^2-\partial_y {\cal A}_x^2)
\right].
\end{eqnarray}


\end{multicols}

\end{document}